\documentstyle[12pt,epsfig]{article}
\begin{document}
\begin{center}  {Critical Behavior in a Cellular Automata 
Animal Disease Transmission Model}
\end{center}
\vspace{.2in}
\begin{center}
P.D. Morley
\\Veridian Corporation\\National Systems Group\\14700 Lee Road
Chantilly, VA 20151\footnote{e-mail address: 
peter.morley@veridian.com}   
\end{center}
\vspace{.2in}  
\begin{center}
Julius Chang\\Strategic Analysis Inc.\\3601 Wilson Blvd\\Suite 500\\Arlington, VA
22201\footnote{e-mail address: jchang@sainc.com}
\end{center}
\begin{abstract}
Using a cellular automata model, we simulate the British Government Policy 
(BGP)
in the 2001 foot and mouth epidemic in Great Britain. When clinical symptoms 
of the
disease appeared on a farm, there is mandatory slaughter (culling) of all livestock on
an infected premise (IP). Those farms that neighbor an IP (contiguous premise, CP), are also culled, aka nearest neighbor interaction. Farms 
where the disease may be prevalent from animal, human, vehicle or airborne 
transmission (dangerous contact, DC), are additionally culled, aka next-to-nearest 
neighbor iteractions and lightning factor. The resulting mathematical model
possesses a phase transition, whereupon if the physical 
disease transmission kernel exceeds a critical value, catastrophic loss of
animals ensues. The non-local disease transport probability can be as low as .01\% per day and the disease can still be in the high mortality phase.
We show that the fundamental equation for sustainable
disease transport is the criticality equation for neutron fission cascade. 
Finally, we calculate that the percentage of culled animals that are
actually healthy is $\approx 30$\%.
\end{abstract}
\vspace{.2in}   
{\it Keywords:} Phase transition; critical behavior; cellular automata; 
animal disease transmission model; foot and mouth disease.
\newpage

\section{Prologue$^{1}$}
On or about 
February 19th, 2001, on a swill feeding farm (the index farm) at Hedon-on-the-Wall in 
the county of Tyne and Wear, type O Pan-Asian strain foot and mouth virus 
disease (FMD) was introduced into the United Kingdom. 
From there the virus spread (probably by wind)
to a sheep farm 7 km away, and then to 96 different locations in Great 
Britain. By the time the epidemic was identified, 43 farms were infected. To
confront the epidemic, the
British Government instituted a policy of culling (killing of animals) to 
eradicate the infection. One year later with over 4 million animals destroyed
(with unquantifiable human misery), Great Britain was declared disease-free.

\section{Introduction}
Foot and mouth disease is an extremely contagious trans-species 
(Bovidae such as cattle, zebus, domestic buffaloes and yaks; sheep, swine, goats, all wild ruminants, and suidae) 
virus that can live temporally outside the host mammal. 
In this paper, we simulate in a mathematical model the 2001 BGP
in controlling and finally eradicating foot and mouth virus in Great Britain. 
One could call the BGP
"risk aversion" in that it relied on no vaccinations and culling of IP
when clinical symptoms appeared. The main interest here is the question of whether the BGP 
risked an unacceptable loss of animals by not knowing that the underlying FMD
mathematical model exists in different phases.

The FMD stochastic disease propagation cellular automata model used in this paper has strong resemblance to the famous
"forest fire" models$^{2}$. In general, probabilistic cellular automata 
models are a modern 
tool for simulations of dynamical systems. They can model complex 
spatio-temporal structures involving cooperative, time-dependent 
phenomena$^{3}$. 
In contrast to the mentioned forest fire diffusion models which have forest regeneration, 
no steady-state is present with the BGP: the disease becomes eradicated (The British Government prohibited the restocking of culled farms - ALL livestock killed - until after the epidemic was officially declared over). Thus,
the only question is whether the peculiar epidemiology of FMD causes the
BGP to incur surprisingly severe losses. How this may happen is illustrated
in Fig.\ 1. 

\vspace{0.5cm}
\begin{figure}[htb]
\begin{center}
\leavevmode {\epsfysize=3cm \epsffile{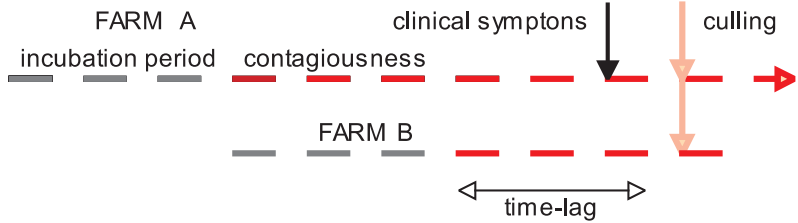}}
\end{center}
\caption[*]{\baselineskip 13pt
Culling destroys the animals of farm A and adjoining farm B, but FMD can have already spread from farm B}
\label{Figure 1}
\end{figure}

There are three time scales:  incubation time, contagious time and 
clinical symptom time. All three time scales depend statistically on the
host type of animal, as well as the particular strain of FMD. For purposes of
illustration, let us choose a 3 day incubation time and a eight day 
clinical
symptom time. Refering to Fig.\ 1, farm A is infected on day 1 and enters its
contagious period on day 4. After day 3 onwards, it can transmit the disease
to neighbors (farm B) with some probablity per day value, which depends on 
the
type of farm B animals, their population and their distance away from farm A. 
On day 8, clinical symptoms appear at farm A, and according to BGP, it 
becomes an IP, with all animals culled (next day). Also, according to BGP, 
farm B is
a CP, and so its animals are culled, as well. As can be seen from Fig.\ 1, 
there is a time period called the "time-lag", whereby the animals at farm B 
may have been contagious
and, before being culled, could have transmitted the disease to yet another
farm. In this illustration, the possible time-lag is 3 days (an extreme case). 
Thus the culling of farms A and B does not necessarily stop the 
diffusion of the disease starting originally from farm A. The existence of the
time-lag is the characteristic epidemiological feature of FMD. In the next 
section, examples
of physical probabilities $\Pi$ are given from the FMD data, and show that the 
disease
has a long spatial tail, making time-lags inevitable when a large number of 
infections are 
present. We
will see that the cellular automata model of BGP with time-lag 
has a phase transition, or
critical point, called
the percolation limit. If the physical 
disease transmission kernel exceeds a critical value, the disease percolates
throughout the landscape, in spite of the culling policy.

Within this paper, three variants of the cellular automata model are 
presented, and discussed in more detail later on. 
All model variants use [1] a 300 $\times$ 300 cellular landscape; [2] a conditional probability per day infection ($\Pi$) which falls off with distance. This cellular automata fall-off distribution is shown in Fig.\ 2.

\vspace{0.5cm}
\begin{figure}[htb]
\begin{center}
\leavevmode {\epsfysize=7cm \epsffile{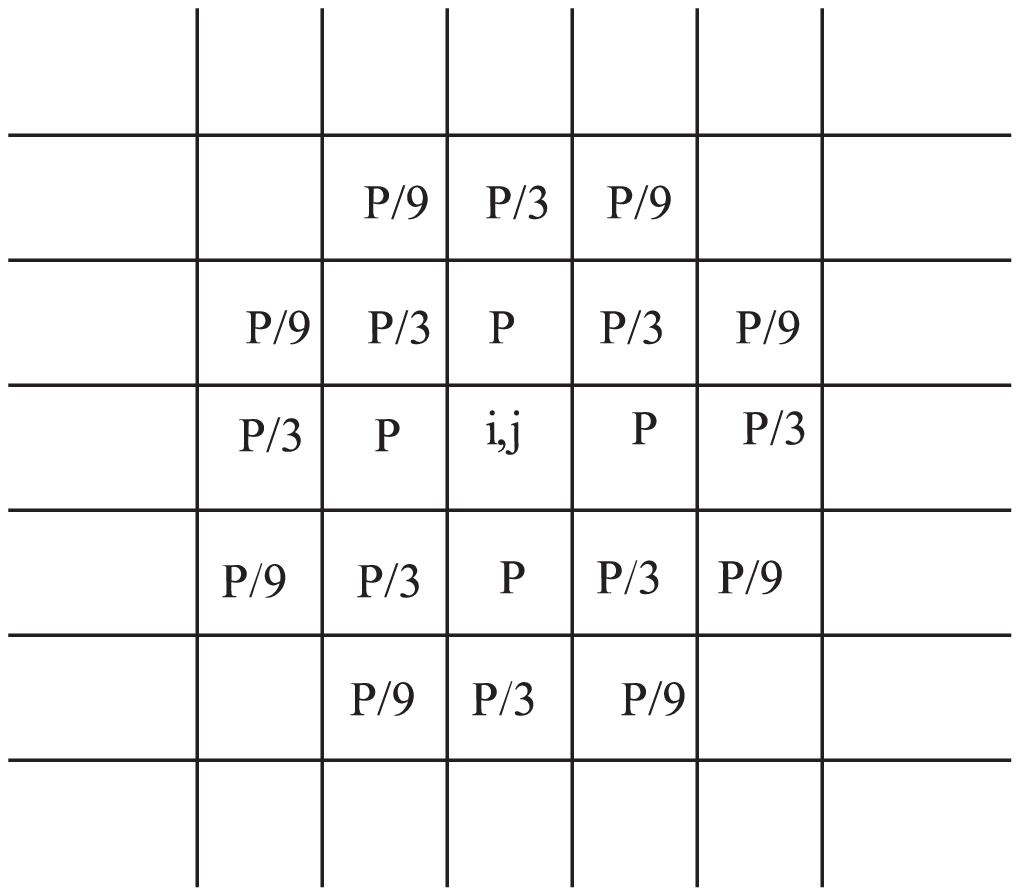}}
\end{center}
\caption[*]{\baselineskip 13pt
Cellular diffusion probability distribution}
\label{Figure 2}
\end{figure}

The middle cell row = i, column = j is the IP. Those cells having a $P$ diffusion probability (discussed below) are the CP. The cells having a $P/3$ diffusion probability are the DC. The infection can jump from an IP to CP, DC or beyond. 

It is important to understand the difference between $P$ and $\Pi$. When an IP is identified in the contagious state, other cells containing healthy animals are at risk from this IP. 
This probability risk increases with the number of healthy animals in the other cell and decreases with the distance from the IP.  Thus $\Pi$ is the conditional probability per day: given that an IP exists X km away, what is the probability per day of infection $\Pi$ if the healthy farm contains N number of animals. The diffusion probability $P$, however, measures the probability per day that a cell will obtain the disease, and so it is equal to the probability that animals are actually present in the cell ($P_{animals}$) times the probability per day that those animals will contact the disease ($\Pi$):

\begin{equation}
P = P_{animals} \times \Pi
\end{equation}

Since England has about 93,000 square miles of area, the $300 \times 300$ lat­tice has a physical
mapping metric of about 1 square mile per cell.
 
The first model variant, called the homogeneous model, has farms in all cells, with a uniform, conditional probability per day of infection from
neigh­bors, $\Pi = {\rm constant}$. That is, an IP can infect
nearest neighbors with a uniform, conditional probability
$\Pi$, if these neighbors still have animals. An IP also
can infect second-nearest neighbor farms with a smaller,
but still uniform, conditional probability $\Pi/3$, and
third-nearest neighbors with a still smaller uniform, conditonal probability $\Pi /9$.
 
The second variant, called the heterogeneous model, adds
a random spatial density of farms (labeled by the
symbol $\nu$, where $\nu$ can range from 0 to 1). We use a
farm density value $\nu = 0.75$. Thus, a cell has a $100 \times \nu$
probability of having a farm. As with the homogeneous
model, an IP can infect nearest neighbors with a uniform,
conditional probability $\Pi$, if these neighbors still have animals,
second-nearest neighbor farms with a uniform, conditional
probability $\Pi /3$; and third-nearest neighbors with a uniform,
conditonal probability $\Pi /9$.
 
The third and most realistic BGP simulation has a heterogeneous
spatial distribution of farms, a stochastic  conditional probability per
day infection of neighbors, $\Pi \neq {\rm constant}$, and, most importantly, a lightning factor (LF $\neq 0$).
The LF represents the non-local transmission probability of FMD$^{4}$. 

\begin{table}[h]
\begin{center}
\begin{tabular}{||c|c|c|c||} \hline
Model Variants & $\Pi$ & Farm density ($\nu$) & LF \\ \hline
homogeneous case & constant & 1 & 0 \\
heterogeneous case & constant & 0.75 & 0 \\
BGP variant & variable & 0.75 & constant ($\neq 0$) \\ \hline
\end{tabular}
\caption{The three models presented}
\end{center}
\end{table}

\section{Conditional Infection Probabilities per Day ($\Pi$)}
The extensive spread of the 2001 FMD in Great Britain allowed for the first 
time identification of the physical transmission probability kernel$^{5}$. 
This kernel is an empirically derived function (depending on the susceptibility factor of each species of animal, the number of animals of each species and the distance from an IP) using FMD data taken directly from the 2001 epidemic. Using this kernel for cattle, we plot in Fig.\ 3 the infection
probability per day for small farms (500 and 1000 cattle) and  
for large farms (2500 and 5000 cattle). This probability per day of infection means that a farm containing N number of healthy animals situated X kms away from an IP, has an empirically derived probability per day of getting the disease. So 
taking the 5000 cattle graph as an example: each day, a large farm containing 5000 healthy cattle has a 60\% chance becoming infected with FMD if an IP is located 6 km away. As already mentioned, we use the symbol $\Pi$ to designate this conditional probability per day attribute associated with a farm containing healthy animals. 

\vspace{0.5cm}
\begin{figure}[htb]
\begin{center}
\leavevmode {\epsfysize=9cm \epsffile{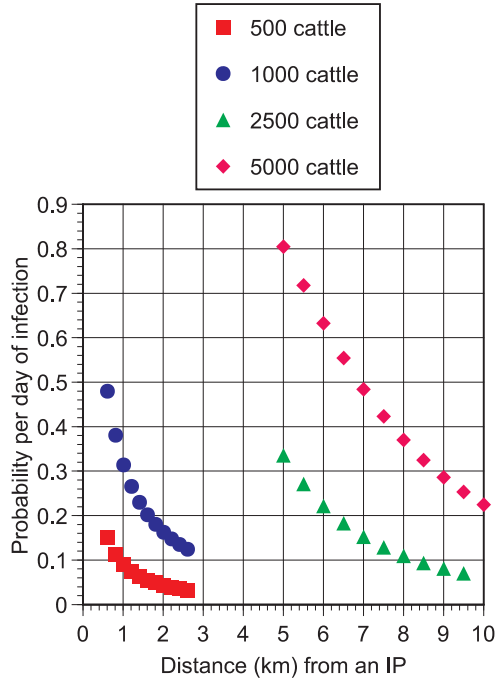}}
\end{center}
\caption[*]{\baselineskip 13pt
Computed conditional probability per day of infection $\Pi$ vs.\ distance (km), based on actual transmission kernel data$^{5}$ from the Great Britain outbreak.}
\label{Figure 3}
\end{figure}

\section{Cellular Automata Model}
The developed cellular automata model is a fully object oriented C++ simulation. The square lattice 300$\times$300 = 90,000 cells (with row i and column j) 
has both time and state information. The possible state information 
is:  Cell[i][j][1] = 0 is culled (originally with animals, but now empty), 
1 is diseased, 2 has unaffected animals, -1 means empty (no original animals). 
Cell[i][j][0] has the time information: On the day of the infection, 
Cell[i][j][0] has the value: -(time-lag)-(incubation); for each 
time unit (a day) 1 is added to this such that when it is -(time-lag), contagious 
transmission can occur and when zero, culling about Cell[i][j] occurs. 
An infected cell remains infected until culling occurs. 
Empty cells are just an impediment to the disease diffusion, not being 
capable of transmitting the disease. One of the more interesting questions that a simulation can address is to compute the percentage of culled animals that are actually healthy. It is relatively simple to keep track of the state of animals at the time of culling, and we find that the percentage healthy can be as high as 40\%, with a typical average of around 30\%. 
 
In Fig.\ 4, we give the logic of the program, while in Fig.\ 2, the individual
diffusion probabilities per day ($P$) are shown, relative to IP Cell[i][j]. 
When IP Cell[i][j] has reached clinical symptoms, all CP are culled (those
with diffusion probability $P$) and all DC are culled (those with diffusion probability $P/3$).

\vspace{0.5cm}
\begin{figure}[htb]
\begin{center}
\leavevmode {\epsfysize=9cm \epsffile{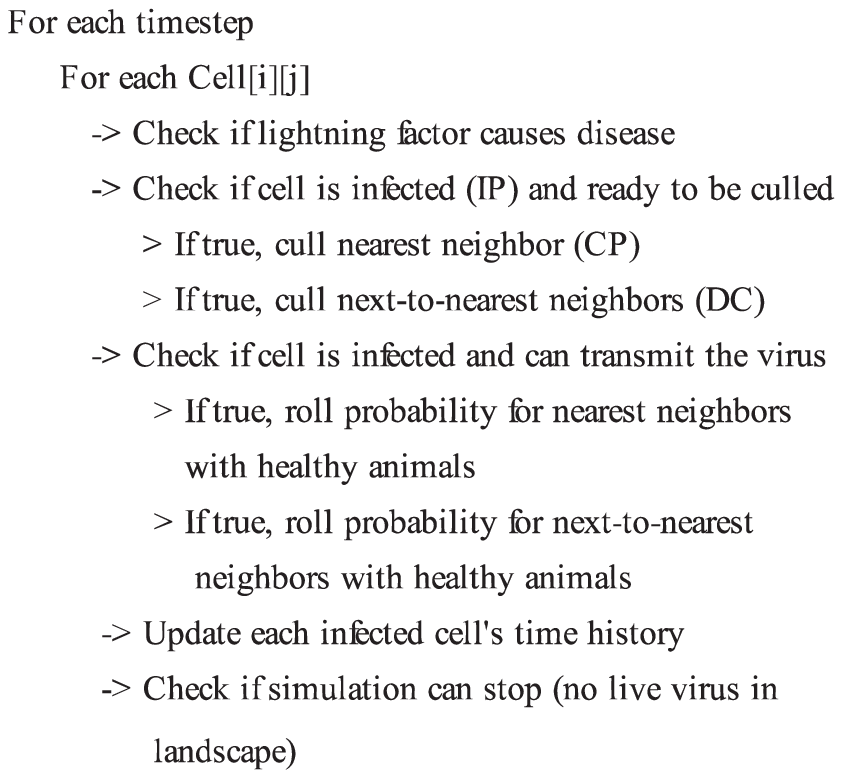}}
\end{center}
\caption[*]{\baselineskip 13pt
Program logic}
\label{Figure 4}
\end{figure}

Since the cellular automata model is a Monte Carlo simulation involving
millions of random numbers, it is important to use an unbiased
random number generator. In this paper, the "gold plated"$^{6}$ ran2 was
used. Each run had different random numbers because the C++ time( ) function
was used to set the value of the random number seed.

As mentioned previously, the main interest in the research is the question of whether the BGP 
risked an unacceptable loss of animals by not knowing that the underlying FMD
mathematical model exists in different phases. Thus we are looking
for critical behavior as we vary the model parameters. There are multiple
ways to detect a phase transition. One is to plot the mortality or survivor
ratio versus the transmission probability per day and see a sharp transition. A second method is to plot the duration time 
of the epidemic versus transmission probability per day
and see that a peak duration appears. The larger the lattice size,
the more pronounced is the phase transition.

Once the existence of the phase transition is established, there are interesting
quantities that can be investigated, such as critical exponents and the 
disease time correlation function. A particularly challenging problem
is to compute analytically the critical probability of the phase transition 
for variants of the model. In actual practice, one needs to go to
extremely large lattices to see power law and fractal scaling behavior, because
small lattices produce a finite width for the phase transition.
The $300 \times 300$
lattice used here, though producing a substantial width to the phase
transition, is more than adequate to demonstrate the underlying
mathematics of the BGP.
 
\subsection{Homogeneous Variant}
The homogeneous version of the model has farms in all cells. An initial,
randomly chosen cell is infected and begins the simulation. 
We chose to mitigate edge effects by inserting the initial infection within (100,250) instead of using periodic boundary conditions, because the former is computationally far simpler while achieving the same end result of avoiding broadening of the phase transition.
In each run of this homogeneous variant, the transmission probability per day 
$\Pi$ is constant. Recall that $\Pi$ is applied only to healthy farms. 500 runs were averaged for each value of $\Pi$. Results
were also obtained using 1000 runs and showed no significant differences
from the 500 batches. The model parameters of time-lag (number of days
between end of incubation and beginning of culling) and incubation were
set to two days and one day respectively. The real FMD would have these as variable.

In Fig.\ 5, the ratio of the number of surviving farms (not culled) 
to the number
(homogeneous case: 90,000) initially present is plotted as a function of $\Pi$. 
A phase transition is indicated. It represents the percolation critical
point whereby the disease has either zero density or approaches 100\% density (as one goes above the critical probability $\Pi_{c}$) in the 
limit
of very large landscape area (limit L $\rightarrow \infty$). In the two-dimensional cellular automata lattice used here, above and very near the critical probability $\Pi_{c}$, the disease density will scale as $(\Pi - \Pi_{c})^{5/36}$; as soon as the conditional probability moves away from this scaling region, the disease density $\rightarrow$ 100\%.
The critical
transmission probability $\Pi_{c}$ for the homogeneous case is $\Pi_{c} \approx .33$. The epidemic time durations versus $\Pi$ are plotted in Fig.\ 6. 
Again, a peak is seen at $\Pi \approx .33$. In order to be certain this is a phase transition, we run the homogeneous variant using a smaller size lattice of $100 \times 100$ cells and a larger size lattice of $700 \times 700$ to compare directly to the $300 \times 300$ case, Fig.\ 7. As expected, the smaller lattice has a pronounced widening and the larger lattice has a pronounced narrowing. We thus have an unmistakable phase transistion.
In terms of the $300 \times 300$ lattice used, this critical probability is reached for an average 
cattle density $\approx 1500$; since pigs have a much higher FMD susceptibility
coefficient, this critical probability is achieved with a much smaller
number of pig population, or with a smaller combined pig and cattle population.

\subsection{Heterogeneous Variant}
The heterogeneous model differs from the previous homogeneous case by randomly selecting some cells to be vacant. As mentioned earlier, vacant cells are an impediment to disease diffusion. Runs were done
on the same 300$\times$300 lattice and the farm occupancy factor, $\nu$, was chosen to be $\nu = 0.75$. In Fig.\ 5 we plot the survival ratio as a function of $\Pi$, while 
in Fig.\ 6, the epidemic duration as a function of $\Pi$ is plotted. Again, a clear signal of a phase transition is present with the critical conditional probability per day transmission of $\Pi_{c} \approx 0.44$ for this heterogeneous case.

\vspace{0.5cm}
\begin{figure}[htb]
\begin{center}
\leavevmode {\epsfysize=9cm \epsffile{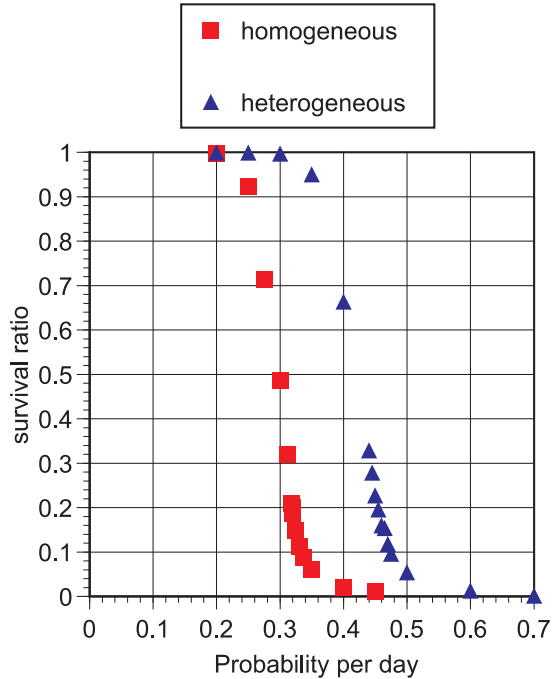}}
\end{center}
\caption[*]{\baselineskip 13pt
Farm survival ratio (fraction of farms not culled)
versus conditional transmission probability per day, $\Pi$, for both homogeneous and heterogeneous cases.}
\label{Figure 5}
\end{figure}

\vspace{0.5cm}
\begin{figure}[htb]
\begin{center}
\leavevmode {\epsfysize=9cm \epsffile{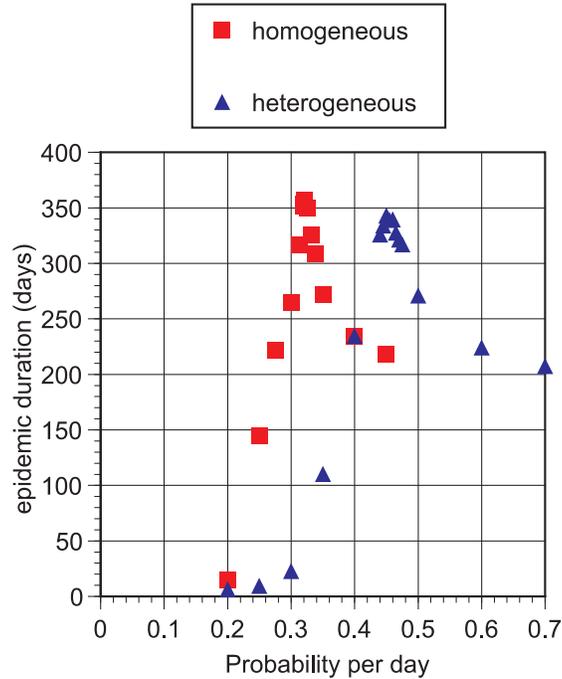}}
\end{center}
\caption[*]{\baselineskip 13pt
Epidemic duration time (in days) versus
transmission conditional probability per day, $\Pi$, for the homogeneous and heterogeneous
cases, showing a peak duration which indicates a phase
transition at a critical transmission probability.}
\label{Figure 6}
\end{figure}

\vspace{0.5cm}
\begin{figure}[htb]
\begin{center}
\leavevmode {\epsfysize=9cm \epsffile{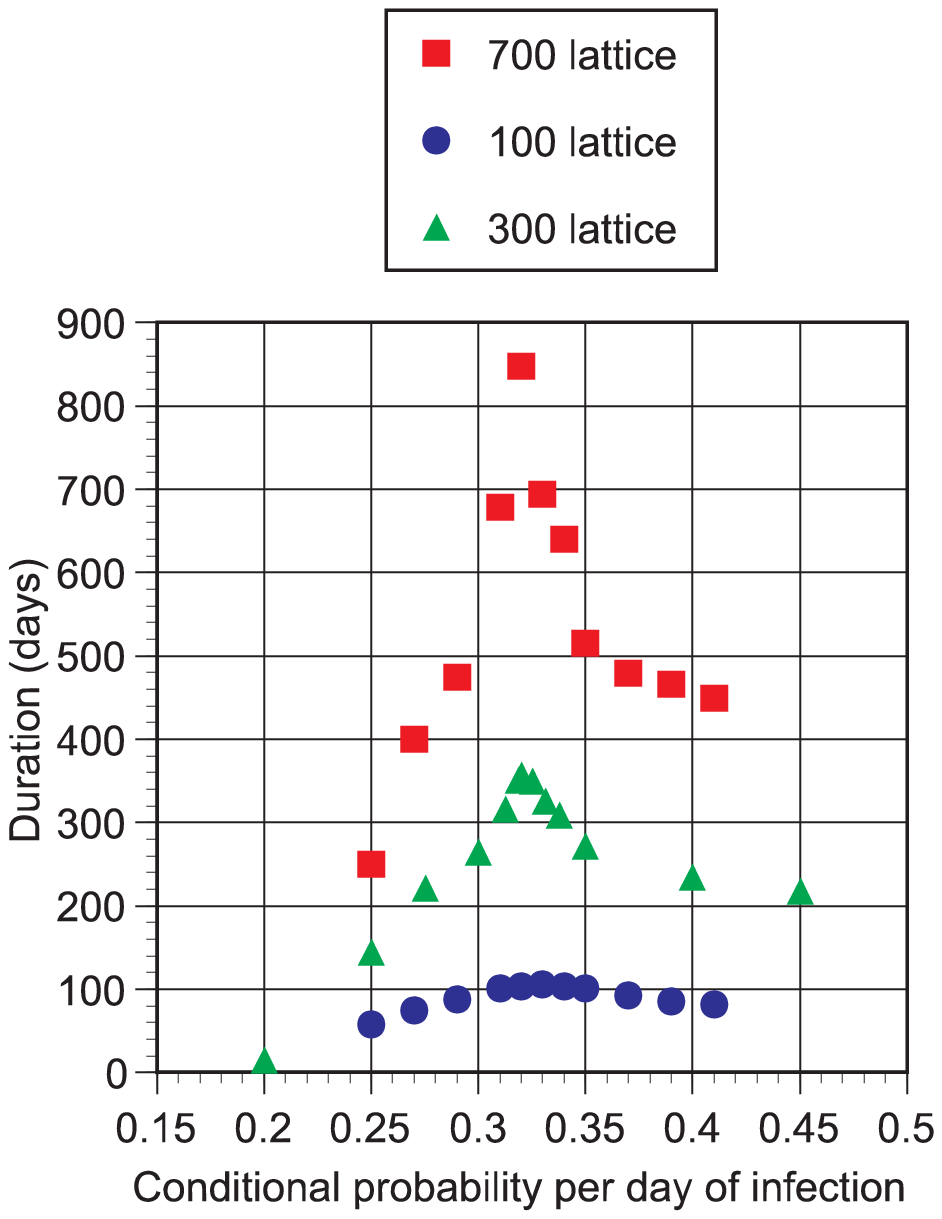}}
\end{center}
\caption[*]{\baselineskip 13pt
Homogeneous epidemic duration time (in days) versus
transmission probability per day, $\Pi$, for the $100 \times 100$ lattice, the $700 \times 700$ lattice and the original $300 \times 300$ lattice}
\label{Figure 7}
\end{figure}

\subsection{Analytical Solutions for Homogeneous and Heterogeneous Variants}
The homogeneous and heterogeneous models are amenable to computation of their
critical conditional probabilities per day. Let us recall$^{3}$ that the square lattice 
has the critical (site) percolation density of 
     \begin{equation}
        P_{percolation} = 0.59273
     \end{equation}
For our situation here, the critical percolation density enters 
into the probability
of finding animals in a cell, allowing the continued diffusion of the 
infection.
Cellular automata models have different neighborhood rules which cause the
critical value of parameters to vary from model to model. In the case here,
the diffusion probability $P$ is a function of $P_{percolation}$. We ask, "What
is the probability of the disease spreading out from IP Cell[i][j] when
nearest neighbor and next-to-nearest neighbors are culled?" 

Refering to Fig.\ 2, all farms with diffusion probability $P$ or $P/3$ are culled, 
so the disease starting out from IP Cell[i][j] must reach cells further 
out within the alloted time-lag.
The time-lag of 2 used here means that if IP farm A has infected a CP or DC,
designated by farm B, on the first day of contagiousness, the disease has 
one additional day to use farm B as a
new source of infection. We first compute the diffusion probability 
that the disease can go outside the culled area.

In the first day of contagiousness, the only cells reachable outside the future
culled area are those that have diffusion probability of $P/9$ and there are 8 of
them, so first day diffusion $P_{first \: day}$ is
    \begin{equation} P_{first \: day} = \frac{8P}{9} \: .
    \end{equation}
However, if the infection successfully infects a CP or DC on day one, it
has a much better chance of escaping outside the culled area on day two. 
On day two, each of the first day's successful jumps is itself a new Cell[i][j], which spreads out according to Fig.\ 2. Thus we need to keep track of the second day's diffusion relative to the future culled cells. So, for example,
if on day one, it reaches a cell with probability $P$, and on the second
day reaches its
neighboring cell of probability $P$, the disease only reaches a future culled cell,
so this mode adds nothing. To achieve diffusion, the second day jump from
farm B must reach at least its $P/3$ cell. The computation of all possible
routes is simple, but tedious. In Table 2, we tabulate the probabilities. The last column is computed by multiplying the first day's probability by the second day's probability and summing over all of the routes that avoid the future culled cells.

\begin{table}[h]
\begin{center}
\begin{tabular}{||c|c|c||} \hline
 first probability & second probability &  total probability \\   \hline
P/9 & \mbox{} & 8P/9 \\
P &  P & 0 \\
P & P/3 & 20P$^{2}$/3 \\
P & P/9 & 16P$^{2}$/9 \\
P/3 & P & 8P$^{2}$/3 \\
P/3 & P/3 & 32P$^{2}$/9 \\
P/3 & P/9 & 48P$^{2}$/27 \\
\mbox{} & P/9 & 8P/9 \\ \hline
\end{tabular}
\caption{Basic probabilities to avoid culled cells}
\end{center}
\end{table}

By adding up all the probabilities and seting the sum to equal one, we define
the meaning of the critical diffusion (percolation) probability $P_{c}$
    \begin{equation} 
    1 = \frac{16P_{c}}{9} + \frac{148 P_{c}^{2}}{9}  \: .
    \end{equation}

Had we used a time-lag of three days, then the resulting polynomial would have been cubic, changing to quartic for a time-lag of four days and so on. 

The infection cannot continue to percolate, however, unless the cell it 
has jumped to actually has animals. These potential jumped cells may have 
already been 
culled in earlier time periods. Futhermore, a likely jumped cell may be
originally vacant, with probability $100 \times(1-\nu)$, as happens in the heterogeneous 
case. The probability $P_{animals}$
that any
of the possible jumped cells has animals at exactly the critical diffusion 
limit is
       \begin{equation}
       P_{animals} = \nu P_{percolation}
       \end{equation}

Thus the critical diffusion probability $P_{c}$ for disease percolation is a product 
of the  
critical transmission probability per day for disease transmission, $\Pi_{c}$, when animals are present,
times the probability that animals are actually present:
     \begin{equation}
     P_{c} = \Pi_{c} \times \nu \times P_{percolation}  \: .
     \end{equation}
Solving Eq.(3) gives
    \begin{equation}
    \Pi_{c} \times \nu \times P_{percolation} = 0.1984 \: .
    \end{equation}
Letting $\sigma_{multiple} = 1/.1984$, where each infection generates a possible
$\sigma_{multiple}$ new ones outside of the culled area, the fundamental equation for disease
percolation becomes
  \begin{equation}
    \Pi_{c} \times \sigma_{multiple}\times \nu \times P_{percolation}
    = 1 
    \end{equation}

The product $\Pi_{c} \times \nu \times P_{percolation}$ is the probability that there are animals in a cell and that they will obtain the disease. The number of cells outside of the future culled area succesfully jumped to is $\sigma_{multiple}$. Their product must be equal to one in order for the disease to be sustainable. 
Those readers familiar with the design of nuclear reactors will see that
Eq.(8) is the neutron equation for criticality.

The analytical solution of $\Pi_{c}$ for the homogeneous and 
heterogeneous cases is given 
in Table 3. For the heterogeneous runs, the constant entering the landscape farm density probability, $\nu$, had the value of 0.75.

Eq.\ (8) explains why there are two phases for the BGP: if the average
diffusion probability is less than that required for escaping the culled
cells, then the infection
cannot be sustained, and damps out; in the limit of
large landscape, the disease density is zero. Unfortunately with FMD, the
long spatial tail and time-lag require an enormously impractical culling area.

\begin{table}[h]
\begin{center}
\begin{tabular}{||c|c||} \hline
homogeneous case & heterogeneous case \\ \hline
0.335 & 0.335/$\nu$ \\ \hline
\end{tabular}
\caption{Analytical critical conditional probabilites per day $\Pi_{c}$}
\end{center}
\end{table}

In variant 3, where a stochastic conditional probability per day is used, 
$\Pi$ is required 
to satisfy the initial condition $\Pi_{LOW} \leq \Pi  \leq \Pi_{MAX}$, for each farm. That is, each Cell[i][j] that happens to contain a farm (probability $100 \times \nu$ that a cell contains a farm) has assigned to it a random $\Pi$, {\em which is the same as saying that the {\em Cell[i][j]} has a random animal population}.

\subsection{BGP Variant}
This variation of the model is the closest to the BGP. For each run, the location of farms on the $300 \times 300$ lattice is random, as in the previous heterogeneous case (again the probability that a cell contains a farm is $100 \times \nu$). In addition, the animal population of each farm varies randomly, but fixed for that entire run. Thus each run represents an idealized realization of England with an initial random location of farms, each with an initial random farm animal population, categorized by a cell's value of $\Pi$. This $\Pi$ initial information is stored in an extra array: Prob[i][j] holds the randomly assigned $\Pi$ for Cell[i][j], if this cell has a farm. When an IP Cell[i][j] is identified, the location of this IP to other cell locations Cell[i$^{\prime}$][j$^{\prime}]$ containing healthy animals is determined, and the probability 
Prob[i$^{\prime}$][j$^{\prime}]$ is then used (refering back to Fig.\ 4, this is the step
"if true, roll probability \ldots").
Thus, for example, if Cell[i][j] is an IP, then 
Prob[i-1][j+1]/3 is used for the probability that a farm containing healthy animals in 
Cell[i-1][j+1]becomes infected.
 
What is there left to vary? FMD has the frightening ability of non-local disease transport, aka the lightning factor. As long as there is one IP on the landscape, there is a lightning factor for every cell containing healthy animals. This lightning factor is a constant probability per day of non-local infection of healthy animals, a purely random effect. Examples of the lightning factor for FMD include airborne plume of virus, and the spread of FMD to locations outside the local area by movement of persons, vehicles and animals through markets and agents. In Table 3, we present the results. The lower value of $\Pi$ (0.05) represents about 500 cattle (0.05 probability for 1 mile distance from an IP), while the higher value of 0.70 represents about 2500 cattle (0.70 probability for 1 mile distance from an IP). 

\begin{table}[h]
\begin{center}
\begin{tabular}{||c|c|c|c|c|c||} \hline
$P_{LOW}$ & $P_{MAX}$ & Survival ratio & Duration (days) & Lightning & Farm density $\nu$ \\ \hline
0.05 & 0.70 & 0.932 & 134 & 0 & 0.75 \\ 
0.05 & 0.70 & .122 & 386 & 0.0001 & 0.75  \\ \hline
\end{tabular}
\caption{Variant 3 Results}
\end{center}
\end{table}

For zero lightning factor, the disease is in the low mortality phase. A lightning factor as low as 0.0001, however, pushes the disease into the high mortality phase. No attempt was made to analytically predict variant 3.                                                                      

\section{Conclusion}
The underlying mathematical model of the 2001 BGP in the FMD epidemic 
possesses a phase transition. 
For a small average disease transmission kernel, the infection is quickly 
eliminated at hotspots and in the limit of a large area, the density of the 
disease is zero. A second outcome occurs if the average disease 
transmission kernel and its associated lightning factor are greater 
than well-defined numerical critical values. This results in the disease 
propagating throughout the landscape and, in the limit of a large area, 
the disease density approaches one as the conditional probability is raised above the critical value. FMD has wreaked havoc. The existence of the 
phase transition comes from the epidemiological feature of FMD: time-lags 
exist for any culling policy relying on the appearance of clinical symptoms. 

In the United States, there are major cattle producing areas (e.g. Florida) which have similar cattle density and spatial farm distribution as England. Thus implementing the BGP would be a great risk to the United States.
The authors suggest the following alternative strategy: it is possible to 
detect the presence of the virus in animals before clinical symptoms or 
antibodies appear. This is done by PCR (polymerase chain reaction: a method to duplicate RNA or DNA and thus amplify the amount of genetic material in the sample) amplification of viral RNA in the 
host's serum. Recently, it has been shown$^{7}$ that the FMD virus can be detected 
this way in animals. This technology means that the time-lag can be 
eliminated if mobile PCR laboratory units can be made to return results 
in a few hours. Stationing of these units at ports-of-entry and within the 
interior will allow rapid first-responder capability. Identifying the virus 
at an IP before the clinical symptoms appear means the culling policy will 
eradicate the disease efficiently: with no time-lag, no landscape 
percolation exists. Coincident with the PCR mobile units is the 
critical requirement 
of creating a FMD predictor model, using a complete database of farm animals. The FMD predictor model would direct the movement of the mobile units to predicted high risk localities. Without such a code, one squanders away the only real assest that will be effective in confronting this most formidable microorganism. With the use of first-responder units, a draconian movement restriction policy is essential to lowering the lightning factor of non-local disease transport. Finally, the United States must
design and manufacture, as quickly as possible, mobile crematoria with high throughput.

\end{document}